\documentclass[11pt,twoside]{article}
\usepackage{graphicx,epsfig,natbib}
\usepackage{CS18}

\newcommand{\ali}{$A$(Li)}
\newcommand{\cratio}{$^{12}$C/$^{13}$C}
\newcommand{\msun}{$M_{\sun}$}

\begin{document}
%
% NTA: Ignore the setcounter comment here
%\setcounter{page}{7}
%
% NTA: Update your full title here
%
\title{Lithium Inventory of 2 \msun\ Red Clump Stars: Is Li Created During the He Flash?}
%
% NTA: enter your author name, affiliation/address information here
%
\author{Joleen K. Carlberg$^{1}$, Katia Cunha$^{2}$, Verne V. Smith$^{3}$}
\affil{$^1$  Department of Terrestrial Magnetism, Carnegie Institution of Washington, 5241 Broad Branch Road, NW, Washington, DC, USA, 20015}
\affil{$^2$Observat\'orio Nacional, Rua General Jos\'e Cristino, 77, 20921-400 S\~ao Crist\'ov\~ao, Rio de Janeiro, RJ, Brazil}
\affil{$^3$National Optical Astronomy Observatory, 950 North Cherry Ave.,Tucson, AZ, USA, 85719}
\begin{abstract}
A recent study of Li abundances in field red giant stars  suggested that the phenomenon of enriched surface Li may be a short-lived phase of red clump evolution for stars with masses near 2~\msun. ÊAlthough the exact mechanism for generating this Li is not fully understood, it may be related to the He-core flash that immediately precedes the red clump stage. To test the incidence and timescale of this proposed process, we are targeting  red clump stars in four southern open clusters, using the cluster ages to ensure that the stellar masses are $\sim$2~\msun. ÊAdditionally, we observe at least one upper red giant branch star in each cluster to establish the baseline Li abundance prior to the He flash. ÊHere, we present preliminary results on the relative abundances of Li in the clusters' red clump stars. 
\end{abstract}
%
%
%
%
% NTA: Here is where the body of your text goes.
%
\section{Introduction: Why are ``Hot'' Li-rich Giants Interesting?}

Red giant stars (RGs) are known to  show depleted levels of Li in their envelopes. Standard stellar evolution models 
predict that the first dredge-up (FDU)  in the early stages of red giant branch (RGB) evolution will reduce the stellar surface abundance of Li by a factor of 60 \citep{1967ApJ...147..624I} by mixing the surface layers with the Li-depleted interior.
For a solar model, this translate to \ali\footnote{\ali = $\log N({\rm Li})-\log N({H})+12$}=1.5~dex during the red giant stage.  Most RGs have much lower \ali\ \citep[e.g.,][]{1989ApJS...71..293B}, and this is well understood because Li depletion  occurs throughout the main sequence lifetime for many stars. The standard models therefore predict the maximum amount of \ali\ expected in RGs, and those stars with \ali$ > 1.5$~dex are considered Li-rich.

The existence of Li-rich RGs can be explained with a nucleosynthetic origin of  Li in some situations. \cite{1971ApJ...164..111C} discovered  that Li can be regenerated and maintained by the
$^3$He($\alpha,\gamma)^7$Be  reaction, which operates at $T>10^7$~K, if the subsequent  $^7$Be$(e^-,\nu)^7$Li decay occurs at $T<3\times10^6$~K.  This lower temperature is necessary to prevent the rapid destruction of  the newly synthesized Li. 
In other words, the first reaction must occur in the presence of a fast mixing mechanism to transport $^7$Be to cool regions of the star before it decays.
\cite{1975ApJ...196..805S} showed that this condition is met in bright M giants more massive than $\sim1.5$~\msun.  The base of these stars' convection zones are hot enough to synthesize $^7$Be, and the convective mixing quickly redistributes the $^7$Be throughout the cooler surface layers.
The condition is also briefly met  for K giants less massive  than $\sim 2.3$~\msun\ when they reach the ``luminosity bump'' stage of stellar evolution. This stage occurs when the outward-progressing H-burning shell reaches the deepest layer mixed by FDU. Lower mass stars create a pocket of $^3$He in the envelope, which is distributed evenly by FDU. At the bump stage, the H-burning shell suddenly encounters a higher abundance of $^3$He.   \cite{2008ApJ...677..581E} demonstrated that the resultant $^3$He burning (only some of which follows the Cameron-Fowler pathway to $^7$Li) creates a local mean molecular weight inversion, and this buoyant layer quickly rises into the convection zone, where it can deposit   nuclear by-products. 

The observational signature of these two models  is the clustering of Li-rich giants at specific locations on the HR diagram: on the asymptotic giant branch (AGB)  and at the luminosity bump \citep[see, e.g.,][]{2000A&A...359..563C}.
However, as  more Li-rich red giants are discovered, outliers to these well-understood Li-regeneration processes are found.  \cite{2011ApJ...730L..12K} in particular argued that an additional Li-enrichment mechanism is needed to explain a population of field Li-rich stars that are too hot to be associated with the luminosity bump but that are also evolved enough that FDU should have completed.  They hypothesized that these stars are red clump (RC) stars (although first-ascent RGs can also occupy the same location on the HR diagram) that have regenerated Li during the He-flash---the runaway ignition of He-burning in low mass stars. They further suggested that this Li-regeneration may only operate for stars more massive than $\sim 1.5$~\msun\ (otherwise the $^3$He reservoirs are too small) but less massive than $\sim 2.3$~\msun\ (otherwise the stars' He cores are non-degenerate and no He flash occurs).

\section{Open Cluster Red Giants}
Open clusters can be used to test the proposed He-flash Li-enrichment mechanism by measuring \ali\ in the  RC members. Because the ages of stars in open clusters are much better constrained than field stars, tighter constraints on masses can be obtained for the RGs in open clusters compared to field RGs. 
Additionally, the relative masses of the RGs in each cluster will be significantly smaller than the absolute uncertainty. Another advantage of using open clusters is that the RGs at other stages of evolution can be used to determine benchmark levels of \ali\ for the cluster. These benchmark stars, such as lower RGB stars and RGB tip stars, should have nearly identical evolutionary histories as the  RC clump stars. Ambiguity of first-ascent versus RC stages can still exist in open clusters, but they are mitigated compared to field star samples.

We selected four open clusters with well-defined red clumps for our study: Collinder 110, NGC 2204, NGC 2506, and NGC 6583. Using ages and metallicities published in the literature combined with \cite{2008A&A...482..883M} stellar evolution models, we estimate that the masses of the RC stars in these clusters range from 1.6~\msun\ (NGC 2204) to 2.1~\msun\ (NGC 6583).
Red clump stars as well as a few non-RC benchmark stars in each cluster were observed at Las Campanas Observatory between Jan~2013 and Jun~2014. We used the MIKE spectrograph on the 6.5~m Clay telescope to attain high resolution ($R\approx$32,000--45,000) spectra covering the full optical band at a signal-to-noise of at least 90 per pixel at 6700~\AA.

\section{Preliminary Results}

Li abundances have not yet been measured for all of the stars in our sample. However, we have made a preliminary measurement of all of the stars' radial velocities and have inspected the region near the strongest Li~I lines at $6708$~\AA.  In this way, we have classified the stars into two groups: those for which the Li~I line is distinguishable from the neighboring Fe~I line at $6707.4$~\AA\ and those for which no Li~I line is detectable. Stars with radial velocities inconsistent with the cluster's heliocentric velocity have been excluded.  

\begin{figure}
\plottwo{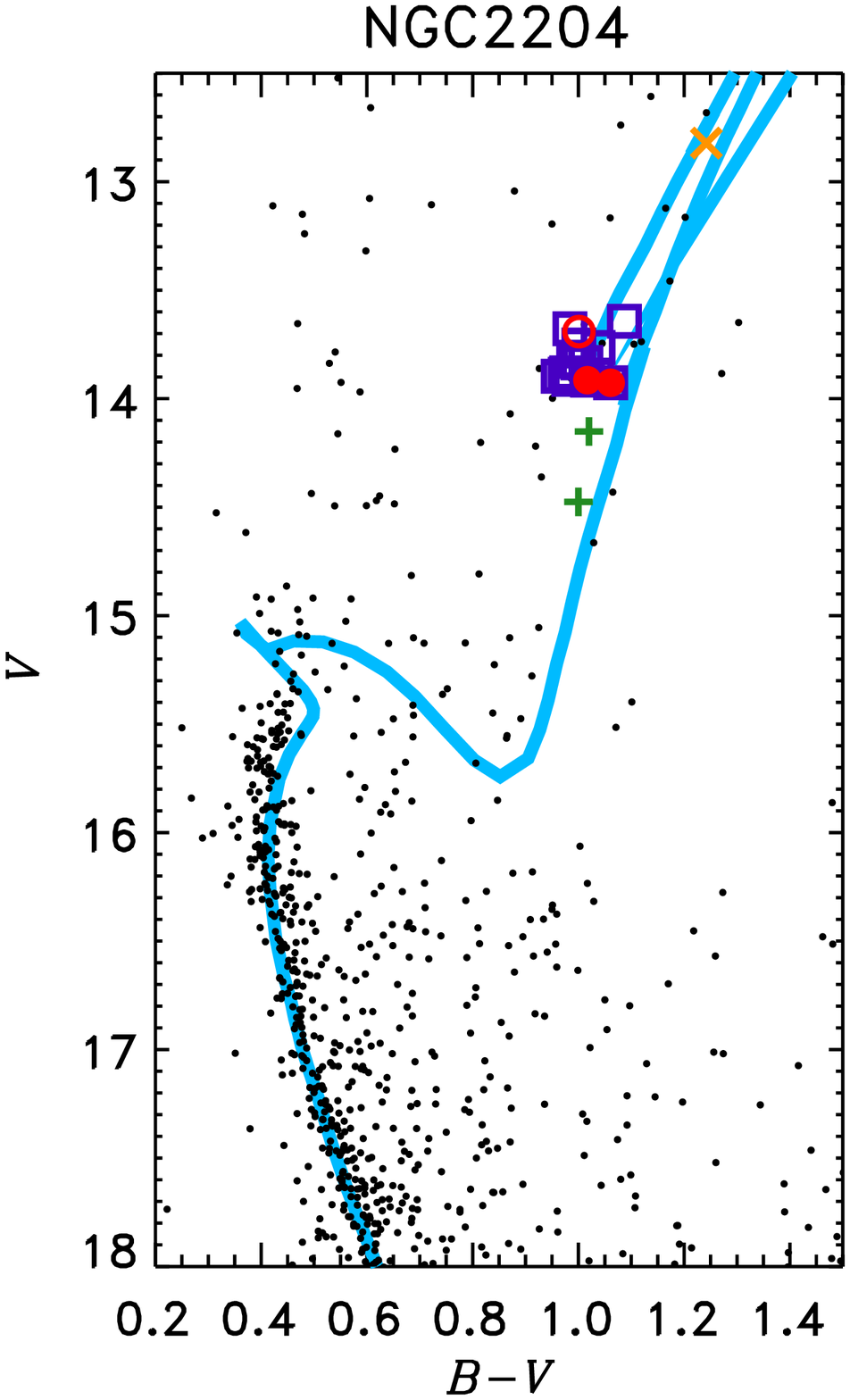}{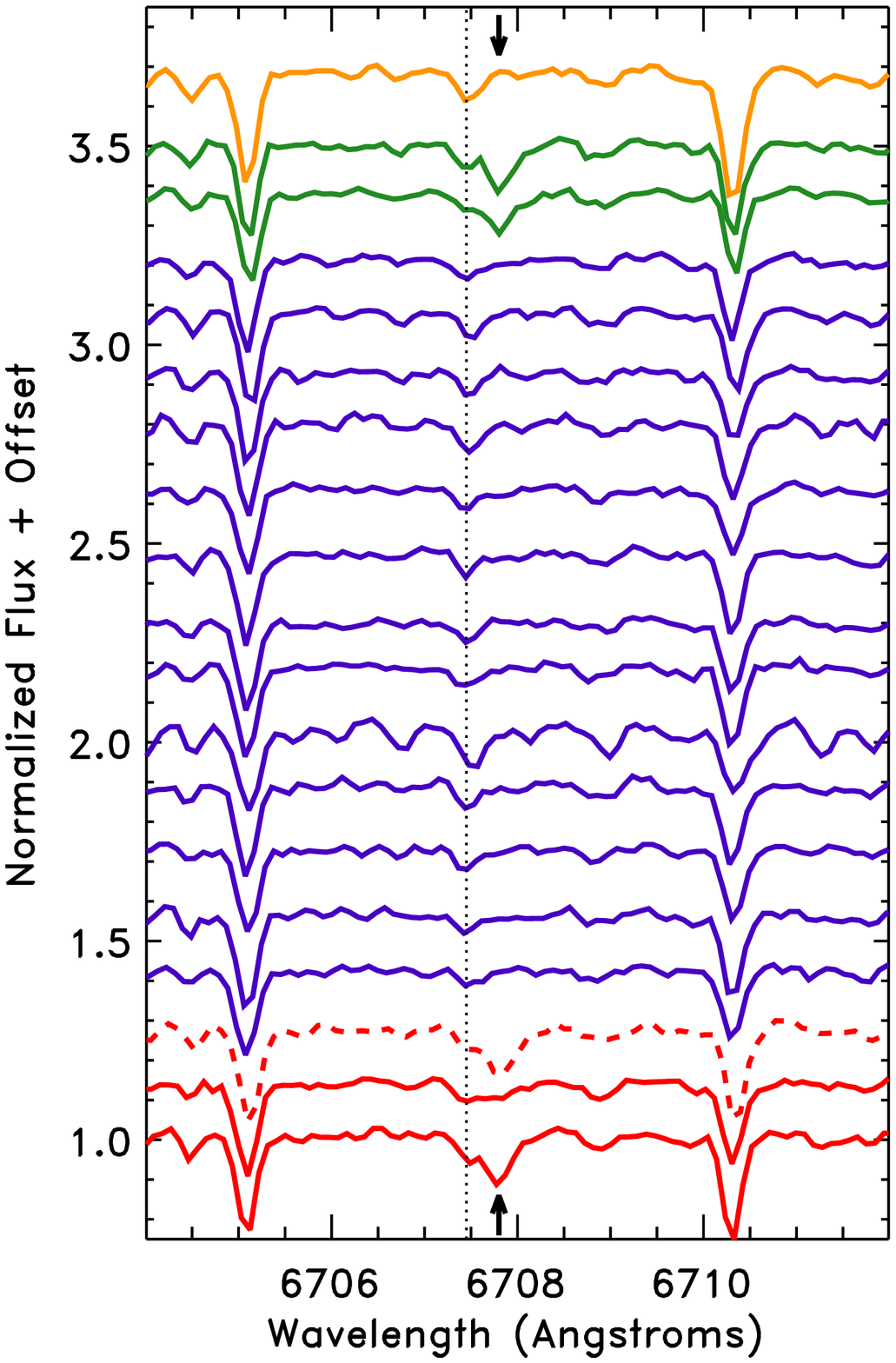}
\caption{CMD (left panel) of NGC 2204 stars together  with a 2 Gyr, [Fe/H]=$-0.23$ isochrone shifted by (m-M)$_0$=13.06 and $E(B-V)=0.08$ \citep{2011AJ....141...58J}. Symbols denote observed RC stars with (red circles) and without (purple squares) visible Li~I lines, RGB tip stars (orange $\times$), and lower RGB stars (green $+$).  The open circle denotes a known spectroscopic binary.  The spectra of the stars (right panel) from bottom to top are the RC stars with and without visible Li~I lines, the lower RGB stars and upper RGB stars.  The dashed line indicates the spectrum of the known binary. The arrows mark the wavelength of the Li~I line, whereas the dotted line shows the location of a neighboring Fe~I line.\label{fig1}}
\end{figure}

\begin{figure}
\plottwo{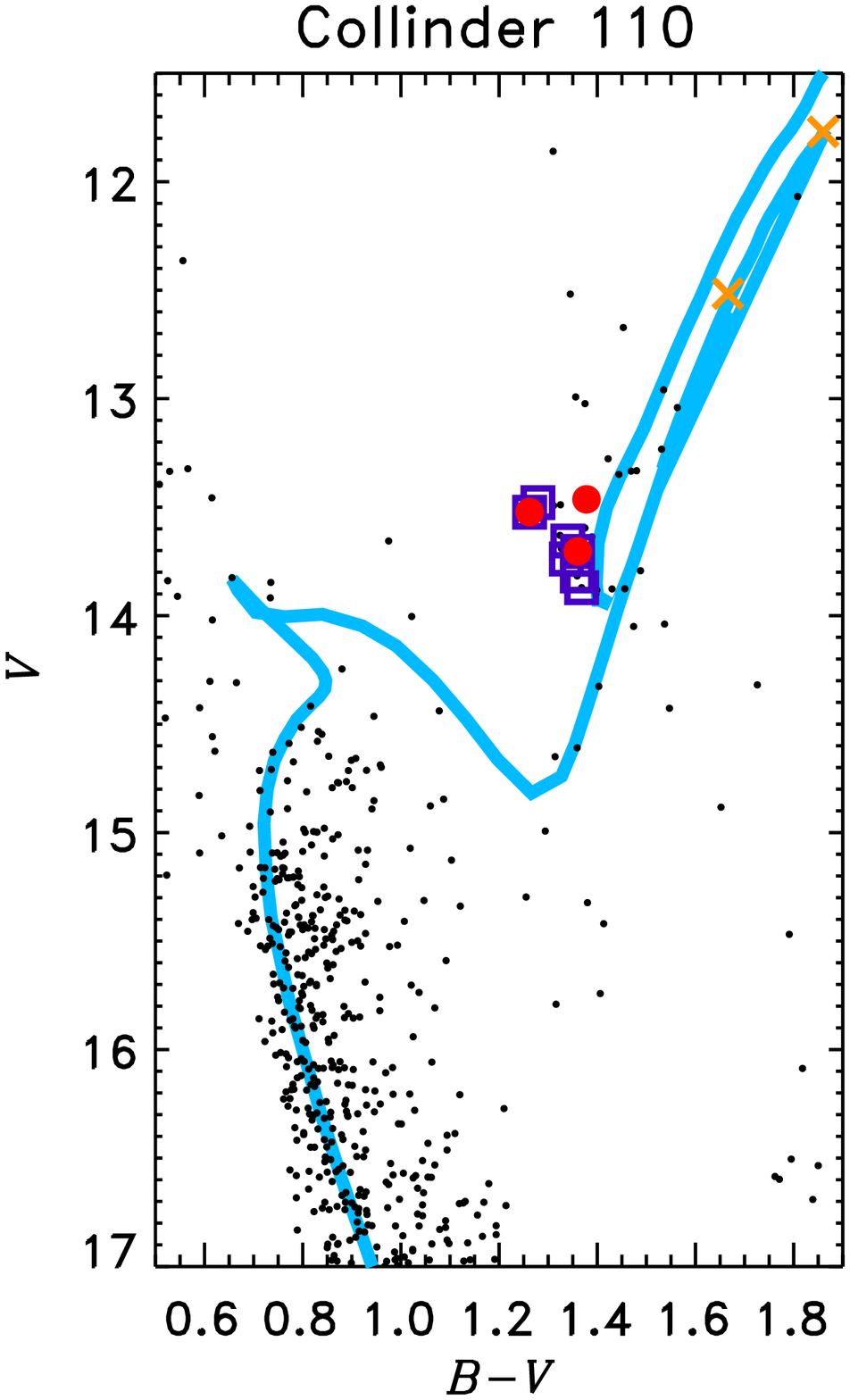}{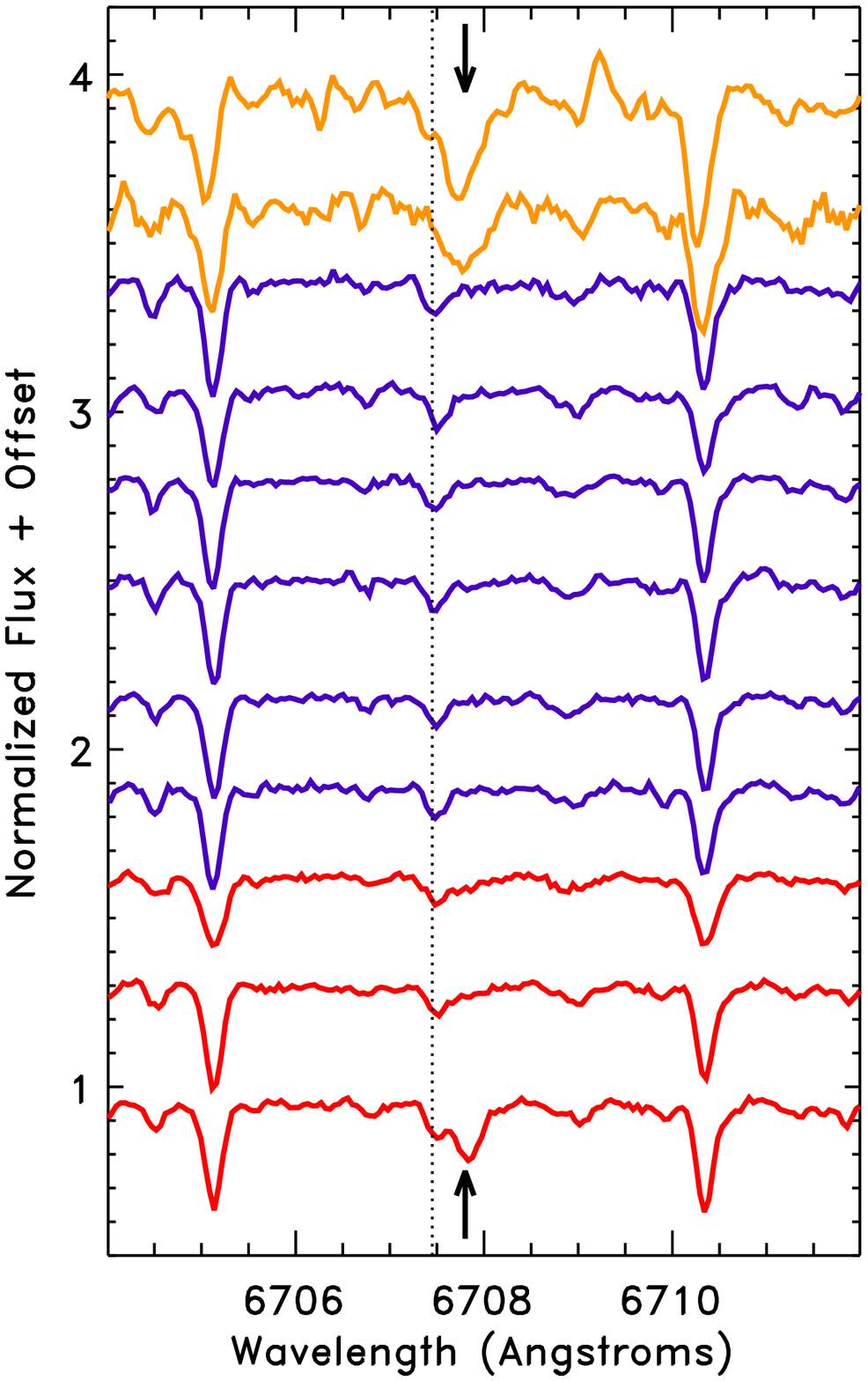}
\caption{ Same as Figure \ref{fig1} for Collinder 110.   The CMD has  a 1.3 Gyr, [Fe/H]=0 isochrone shifted by (m-M)$_0$=11.6 and $E(B-V)=0.38$ \citep{2003MNRAS.343..306B,2010A&A...511A..56P}.    No lower RGB benchmarks were observed for this cluster. \label{fig2}}
\end{figure}

\begin{figure}
\plottwo{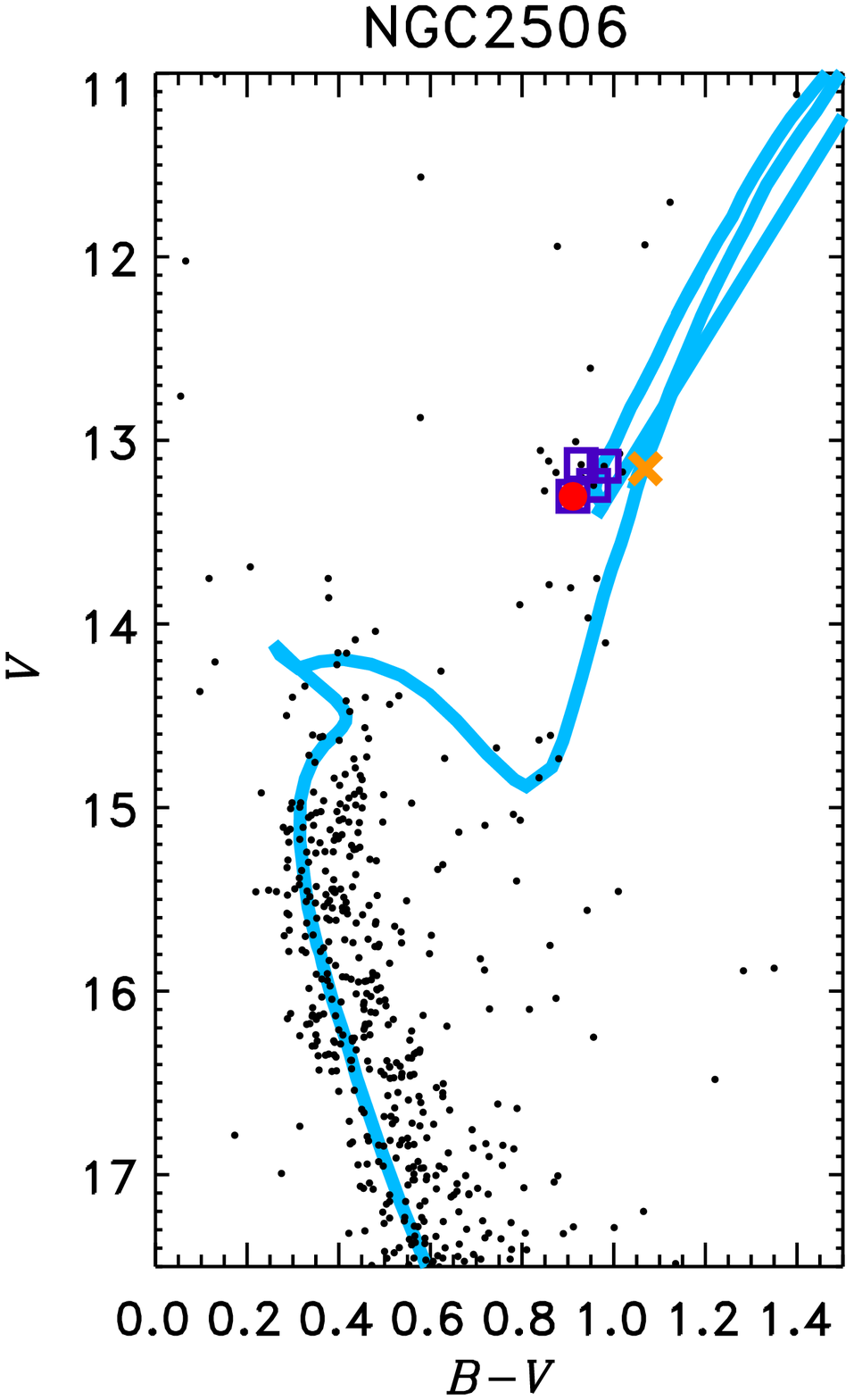}{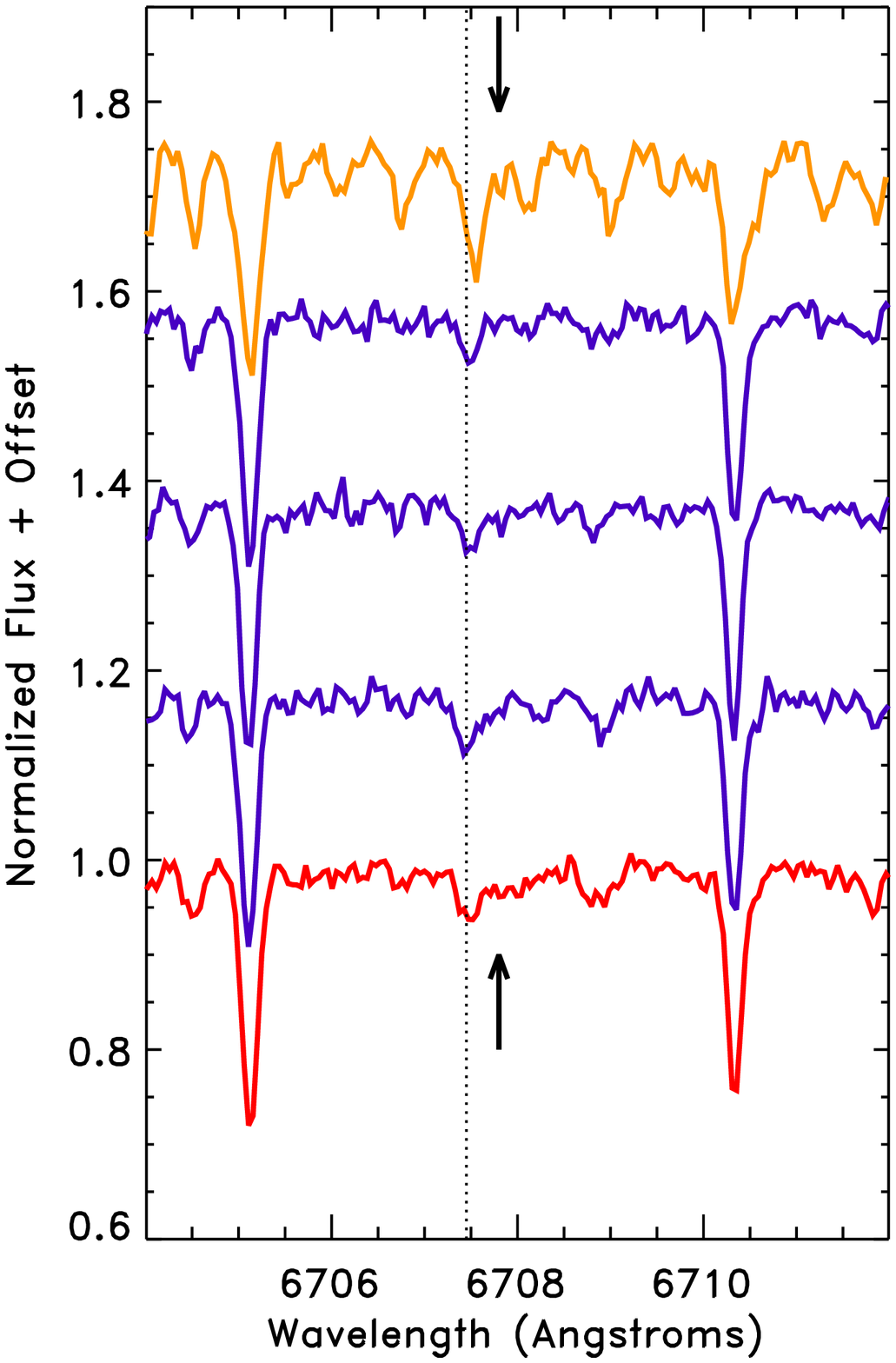}
\caption{ Same as Figure \ref{fig1} for NGC 2506.  The CMD has  a 1.8 Gyr, [Fe/H]=$-0.37$ isochrone shifted by (m-M)$_0$=12.6 and $E(B-V)=0.04$ \citep{1997MNRAS.291..763M}.   The only observed benchmark star is near the luminosity bump.   \label{fig3}}
\end{figure}

\begin{figure}
\plottwo{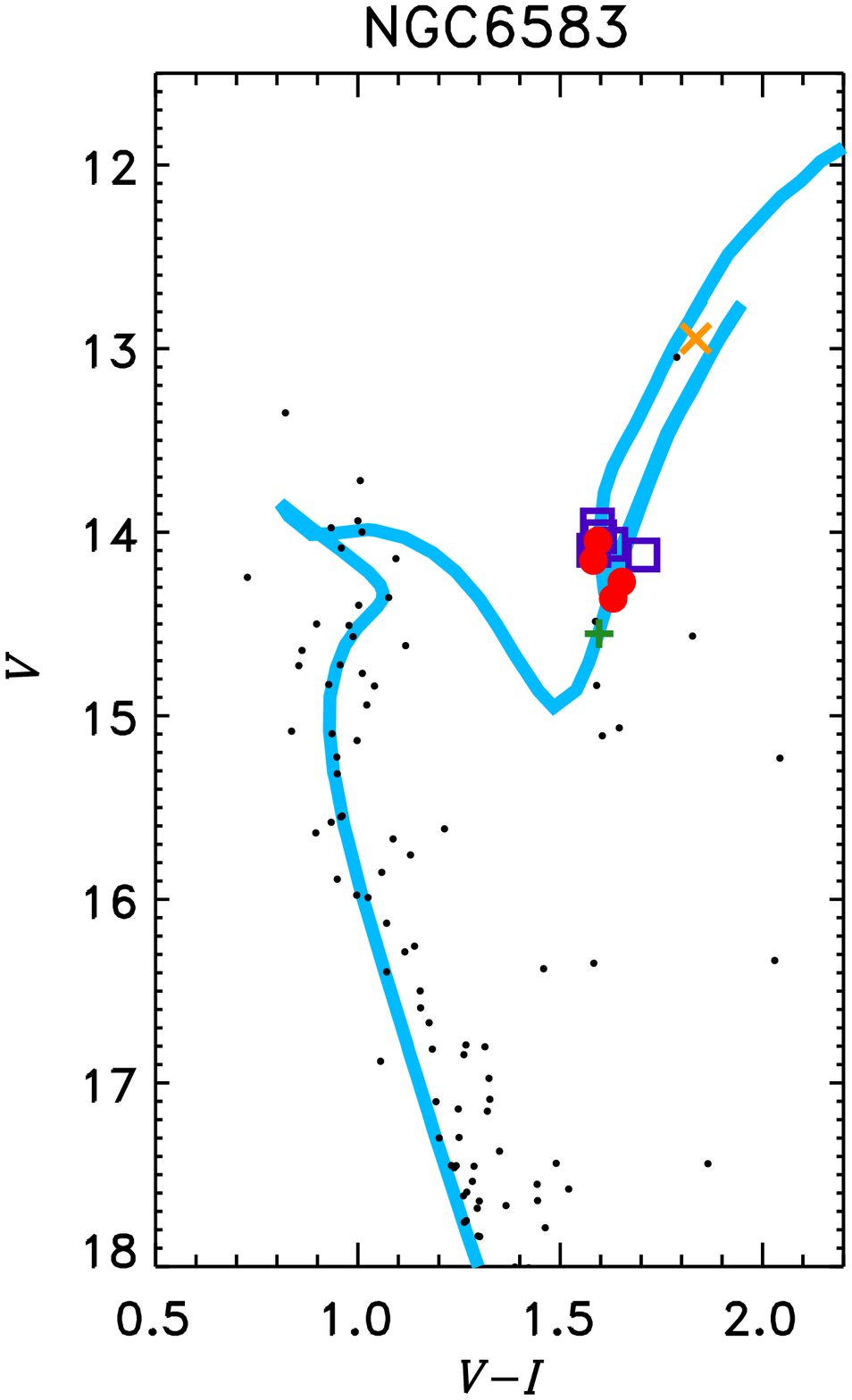}{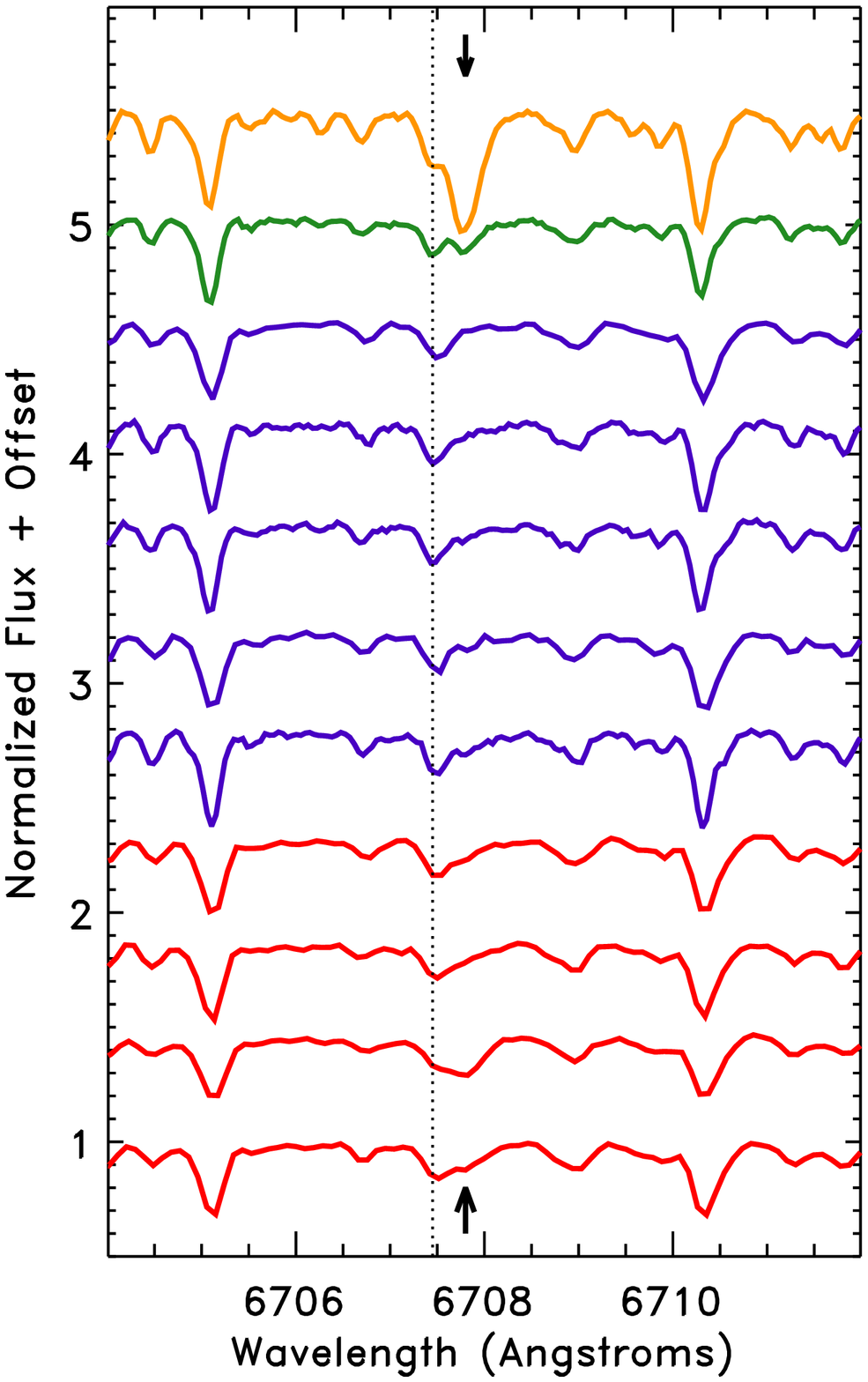}
\caption{ Same as Figure \ref{fig1} for NGC 6583.   The CMD has  a 1 Gyr, [Fe/H]$\sim$0 isochrone shifted by (m-M)$_0$=11.55 and $E(V-I)=0.63$ \citep{2005MNRAS.356..647C}.   \label{fig4}}
\end{figure}

Our preliminary results are summarized in Figures \ref{fig1}--\ref{fig4}, which show  the color-magnitude diagrams (CMDs) and the spectral region around the Li~I line for all of our cluster stars.  The points on the CMD and the spectra are color-coded to indicate whether the stars is a RC star with a visible Li~I line (red), a RC star with no visible Li~I line (purple), a benchmark star at the RGB tip (orange) or a lower RGB star (green).
We find that all of the clusters show at least one candidate RC star with a detectable Li~I line, and in the entire cluster sample 11 of the 38  candidate RC stars (29\%) have detectable Li~I lines.

The behavior of the benchmark stars is more complicated than we expected. 
All three of the lower RGB benchmark stars (green spectra) show detectable Li~I lines, and these stars are in the clusters with both the highest and lowest mass RC stars in our sample.  This suggests that stars in the mass range probed by  our cluster sample are not as severely depleted in Li as  the average RG in the field. This result is perhaps unsurprising since the field RG  population likely probes more lower mass stars, which deplete more Li on the main sequence.
Three clusters have benchmark stars on the upper RGB.  In two of the clusters, these upper RGB  stars have detectable Li~I lines and in the third cluster, they do not.
This difference also appears to be related to stellar mass.  The clusters with detectable Li in the upper RGB are the two clusters with the highest  RC masses, $\sim2.0$--2.1~\msun.

Because the strength of the Li~I line is strongly temperature dependent, becoming stronger at lower stellar effective temperatures, it is difficult to trace the evolution of \ali\ with evolutionary stage in the clusters by eye.
In particular,  we cannot yet comment on the pre-RC Li abundances in the clusters with detectable Li~I lines in the upper RGB stars. However, clusters with upper RGB stars with no visible Li~I line (e.g., NGC 2204) is evidence of severe Li depletion preceding the RC stage. Therefore, we can conclude that detectable Li~I lines in RC stars in this cluster are significant. 
There is some ambiguity in the true evolutionary stages of both the candidate RC stars and some of the benchmark stars that will affect the interpretation of our results. 
Measuring abundances of elements and/or isotopes that   probe mixing (such as \cratio\ in RC stars or s-process elements in potential AGB stars) can help break these degeneracies.

\acknowledgments{ We thank the organizing committee of ``Cool Stars 18'' for all of their hard work in putting together a terrific conference and for the opportunity to present our work in these proceedings.}

\normalsize

\begin{references}

%
% NTA: Instructions on building up your references.
%
% You can get this \bibitem reference format automatically from the NASA ADS Abstract service
% Go to any given reference, eg. http://adsabs.harvard.edu/abs/2013ApJ...775...45V
% and select "Preferred format for this abstract"
% You can set this preferred format under "Preferences" to the "Default format" for the "Reference Format"
%
% You will need to alphabetize these references yourself.
%
% In a perfect world we'd use BibTeX but that's a bit too much of a headache with multiple authors.
%

\bibitem[Bragaglia \& Tosi(2003)]{2003MNRAS.343..306B} Bragaglia, A., \& Tosi, M.\ 2003, \mnras, 343, 306 

\bibitem[Brown et al.(1989)]{1989ApJS...71..293B} Brown, J.~A., Sneden, C., 
Lambert, D.~L., \& Dutchover, E., Jr.\ 1989, \apjs, 71, 293 

\bibitem[Cameron \& Fowler(1971)]{1971ApJ...164..111C} Cameron, A.~G.~W., \& Fowler, W.~A.\ 1971, \apj, 164, 111 

\bibitem[Carraro et al.(2005)]{2005MNRAS.356..647C} Carraro, G., 
M{\'e}ndez, R.~A., \& Costa, E.\ 2005, \mnras, 356, 647 

\bibitem[Charbonnel \& Balachandran(2000)]{2000A&A...359..563C} Charbonnel, C., \& Balachandran, S.~C.\ 2000, \aap, 359, 563 

\bibitem[Eggleton et al.(2008)]{2008ApJ...677..581E} Eggleton, P.~P., 
Dearborn, D.~S.~P., \& Lattanzio, J.~C.\ 2008, \apj, 677, 581 

\bibitem[Iben(1967)]{1967ApJ...147..624I} Iben, I., Jr.\ 1967, \apj, 147, 624 

\bibitem[Jacobson et al.(2011)]{2011AJ....141...58J} Jacobson, H.~R., 
Friel, E.~D., \& Pilachowski, C.~A.\ 2011, \aj, 141, 58 

\bibitem[Kumar et al.(2011)]{2011ApJ...730L..12K} Kumar, Y.~B., Reddy, 
B.~E., \& Lambert, D.~L.\ 2011, \apjl, 730, L12 

\bibitem[Marconi et al.(1997)]{1997MNRAS.291..763M} Marconi, G., Hamilton, 
D., Tosi, M., \& Bragaglia, A.\ 1997, \mnras, 291, 763 

\bibitem[Marigo et al.(2008)]{2008A&A...482..883M} Marigo, P., Girardi, L., Bressan, A., et al.\ 2008, \aap, 482, 883 


\bibitem[Pancino et 
al.(2010)]{2010A&A...511A..56P} Pancino, E., Carrera, R., Rossetti, E., \& Gallart, C.\ 2010, \aap, 511, A56 

\bibitem[Scalo et al.(1975)]{1975ApJ...196..805S} Scalo, J.~M., Despain, 
K.~H., \& Ulrich, R.~K.\ 1975, \apj, 196, 805 



\end{references}
\end{document}